\documentclass{aastex}
\usepackage{emulateapj5}
\usepackage{epsfig}

\newcommand{\firstcite}{(Genzel et al. 1990; Ho et al. 1991; Serabyn, Lacy, \& Achtermann 1992; Zylka et al. 1999; McGary, Coil, \& Ho 2001, hereafter MCH)}

\newcommand{\bq}{\begin{equation}}
\newcommand{\eq}{\end{equation}}

\newcommand{\3}{$_3$}
\newcommand{\kms}{km~s$^{-1}$}

\newcommand{\simgt}{\lower.5ex\hbox{$\; \buildrel > \over \sim \;$}}
\newcommand{\simlt}{\lower.5ex\hbox{$\; \buildrel < \over \sim \;$}}

\begin{document}

\title{Hot Molecular Gas in the Galactic Center}
 
\author{
Robeson M. Herrnstein\altaffilmark{1} and
Paul T.P. Ho\altaffilmark{1}}
\slugcomment{Accepted for publication in ApJ Letters}

\altaffiltext{1}{Harvard-Smithsonian Center for Astrophysics, 60 Garden Street, Cambridge, MA 02138, \\rmcgary@cfa.harvard.edu, pho@cfa.harvard.edu}

\begin{abstract}
Using the new 23~GHz receivers at the Very Large Array (VLA), we have
detected NH\3(6,6) emission ($\nu=25.056025$~GHz) from hot ($>150$~K)
molecular clouds in the central 10~pc of the Galaxy.  This is the
first successful detection of NH\3(6,6) with the VLA.  The brightest
emission comes from a region interior to the ``circumnuclear disk''
(CND), less than 1.5~pc in projected distance from Sgr A*.  This
region does not show molecular emission from lower energy transitions
such as NH\3(1,1) and (2,2), HCN(1-0) and HCO$^+$(1-0).  Line ratios
of NH\3(6,6) and (3,3) emission as well as NH\3(6,6) line widths have
peak values within 1.5~pc of Sgr A*, indicating that the gas is
physically close to the nucleus.  NH\3(6,6) is also detected towards
many features outside the CND observed in NH\3(1,1), (2,2), and (3,3).
These features tend to lie along ridges of gas associated with Sgr A
East or the massive ``molecular ridge'' that connects the ``20~\kms''
and ``50~\kms'' giant molecular clouds (GMCs).  

\end{abstract}


\section{Introduction \label{intro}}
At a distance of only $8.0\pm0.5$~kpc \citep{rei93}, the Galactic
center provides a unique opportunity to study in detail the
environment around a supermassive black hole.  It is now generally
accepted that a black hole of $2.6\times10^6$~$M_\odot$ is located at
the dynamical center of the Galaxy \citep{eck97,ghe98}. In the radio,
emission from just outside the black hole is observed as the strong
($\sim1$~Jy) source, Sgr~A*.  Sgr A* is surrounded by arcs of ionized
gas (Sgr A West) that appear to be feeding the nucleus
\citep{lo83,rob93}.  These arcs are, in turn, surrounded by an
apparent ``ring'' of molecular material called the circumnuclear disk
(CND, \citet{gus87}).

Sgr A West and the CND are located in front of or just inside the
front edge of the expanding supernova remnant (SNR), Sgr A East
\citep{ped89}.  The expansion of Sgr A East appears to be moving large
amounts of material away from the nucleus, forming ridges of material
on all sides \firstcite.  However, some material may move towards the
nucleus after being disrupted by the passing front and three
filamentary ``streamers'' possibly feeding the nucleus have been
detected in NH\3 (Okumura et al. 1989; Ho et al. 1991; Coil \& Ho
1999, 2000; MCH).

The metastable (J=K) NH\3(J,K) rotation inversion transitions at
$\sim23$ GHz have proven to be useful probes of dense (10$^4$--10$^5$
cm$^{-3}$) molecular material near the Galactic center.  They tend to
have a low optical depth and a high excitation temperature at the
Galactic center, making them almost impervious to absorption effects.
Satellite hyperfine lines separated by 10--30~\kms ~on either side of
the main line enable a direct calculation of the optical depth of the
NH\3 emission and line ratios of different transitions can be used to
calculate the rotational temperature, $T_R$, of the gas.

We recently observed NH\3(1,1), (2,2) and (3,3) emission from the
central 10~pc of the Galaxy.
An important result from these studies is the apparent increase in
line width and $T_R$ as gas approaches the nucleus (MCH; McGary \& Ho
in press).  However, the reality of this effect has remained in doubt
because the emission also becomes fainter near Sgr A*.  We suspected
that even the NH\3(3,3) line becomes less sensitive to the extreme
environment near Sgr A* and observed the central 4$'$ (10~pc) of the
Galaxy in NH\3(6,6), at 412~K above ground, using the new 23~GHz
receivers at the Very Large Array\footnote{The National Radio
Astronomy Observatory is a facility of the National Science Foundation
operated under cooperative agreement by Associated Universities, Inc.}
(VLA) in order to detect the hottest molecular gas.  These data
represent the first successful observations of NH\3(6,6) with the
VLA. NH\3(6,6) is detected in many of the features seen in lower NH\3
transitions, but the velocity integrated map is dominated by emission
less than 1.5~pc (40$''$) in projected distance from the nucleus.
Line ratios and extremely broad line widths (50--90~\kms) indicate
this molecular gas is physically close to the supermassive black hole.
The remaining features lie predominantly along the edge of Sgr A East
and have line widths greater than 20~\kms ~indicating interaction with
the shell.

\section{Observations \label{obs}}
Observations of NH\3(6,6) were made with the VLA in the D north-C
array on 2001 October 1 and November 16 and have an identical setup
and spatial coverage to our previous NH\3(1,1), (2,2), and (3,3)
observations so that comparisons between images are straightforward
(MCH).  A five-pointing mosaic centered on Sgr A*
($\alpha_{2000}=17^h45^m40^s.0$, $\delta_{2000}=-29^\circ00'26''.6$)
fully samples the central 4$'$ (10~pc) of the Galaxy and a velocity
coverage of $\pm120$~\kms ~with a resolution of $9.3$~\kms ~includes
most of the velocities observed near the nucleus (the features at
--185 \kms ~observed by \citet{zha95} are not covered).  The data have
been calibrated and images made using the same method as MCH.  The
final beam size after application of a Gaussian taper (FWHM=$10''$) to
the $uv$ data is $12.0''\times9.2''$ with a PA of $-1.52^\circ$.

Observations at $25$~GHz have only recently become possible after the
upgrade of the 23~GHz receivers at the VLA.  
At the time of our observations, 16 antennas were fitted with new
receivers.  At the low declination of Sgr A*, the new receivers have
$\langle T_{sys}\rangle=96$~K, which is comparable to the noise at
23~GHz with the old receivers.  Data are weighted by $1/T_{sys}^2$ so
that old receivers do not increase the noise.  One hour of integration
time was obtained for each pointing resulting in an rms noise per
channel of $\sigma_{ch}=8.7$~mJy~beam$^{-1}$.  For the velocity
integrated image, the $1\sigma$ noise level is 0.23
Jy~beam$^{-1}$~\kms ~and is calculated in the same way as MCH assuming
line emission typically appears in seven channels.

\section{Results \label{res}}

\begin{figure}
\plotone{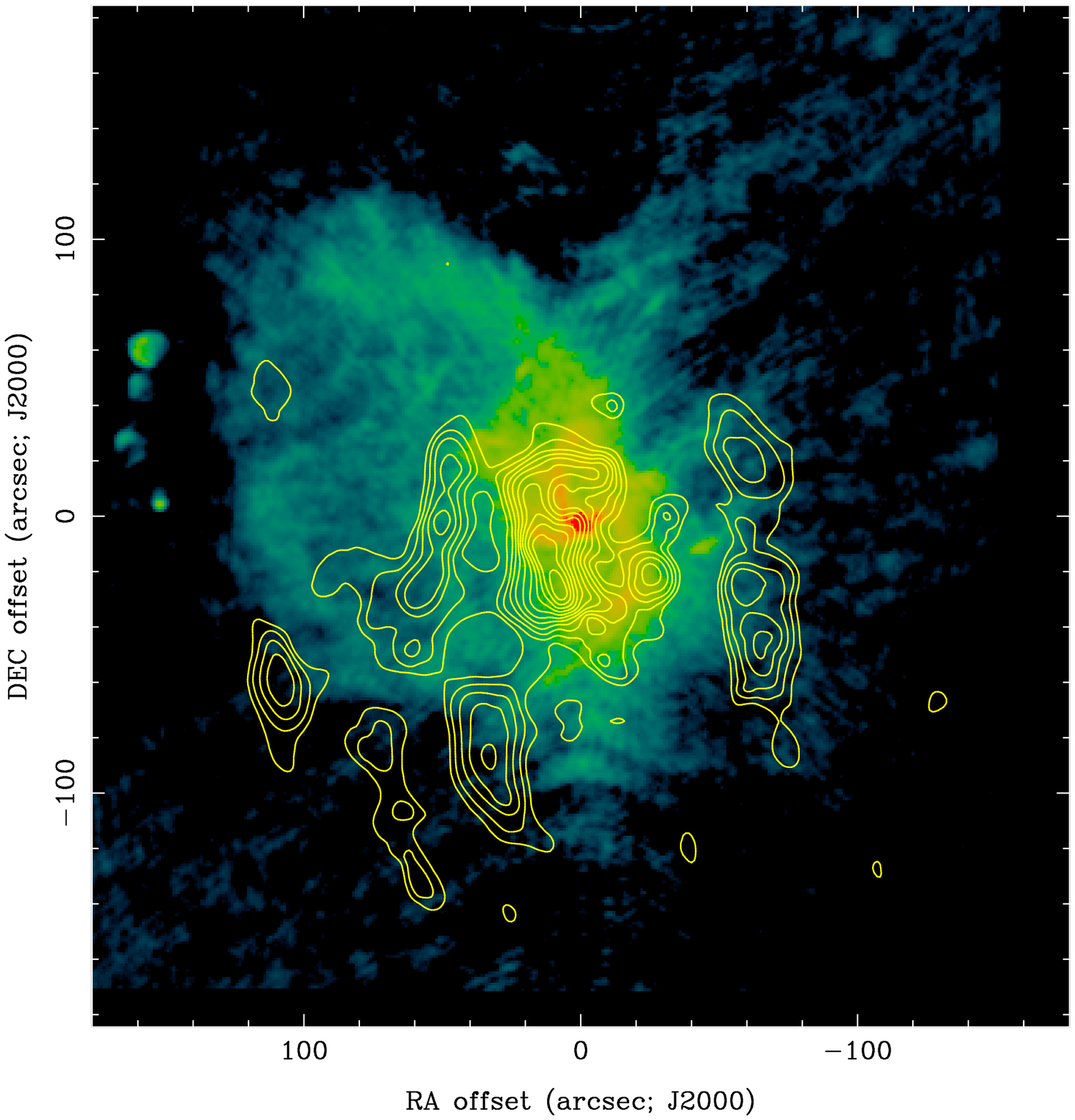}
\caption{NH\3(6,6) velocity integrated emission in contours overlaid
on 6~cm continuum emission in color \citep{yus87}.  Contours are in
steps of 3$\sigma$ where
1$\sigma=0.23$~Jy~beam$^{-1}$~\kms. \label{mom0.fig}}
\end{figure}

\begin{figure}
\plotone{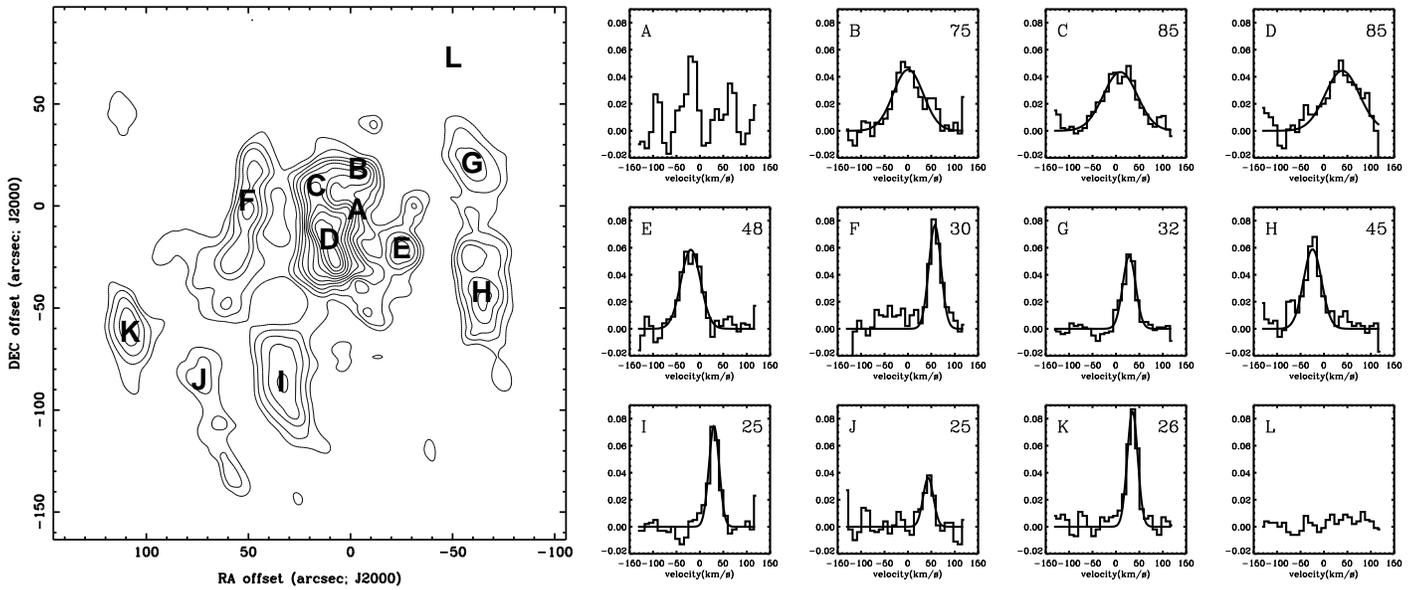}
\caption{$Left:$ Position of spectra overlaid on velocity integrated
NH\3(6,6) emission. Sgr A* is located at (0,0).  $Right:$ NH\3(6,6)
spectra with Gaussian fits overlaid showing that the largest
line widths are located close to the nucleus.  Line widths in \kms ~are
given for spectra B-K in the upper right-hand corner of each
panel. \label{spec.fig}}
\end{figure}

Figure \ref{mom0.fig} shows velocity integrated NH\3(6,6) emission in
contours overlaid on a pseudo-color 6~cm continuum image
\citep{yus87}.  The point source Sgr A* ($\Delta\alpha=0''$,
$\Delta\delta=0''$) is the brightest feature in the 6~cm image.  The
arcs of Sgr A West can be seen in yellow and the large expanding
shell, Sgr A East, is seen in blue.  The contours for NH\3(6,6) are in
steps of $3\sigma$.  Figure \ref{spec.fig} shows the velocity
integrated map with labeled positions for twelve spectra shown to the
right.  The spectra have been Hanning smoothed to a resolution of
$\sim20$~\kms. Spectrum L shows the noise at a position with no
velocity integrated emission.  Note the presence of many separate
features at the position of Sgr A* (spectrum A).  These features are
due to separate clouds, and not satellite hyperfine lines which are
only $\sim30$~\kms ~from the main line. With the exception of the
70~\kms ~feature, the emission is not coincident with the velocities
of known clouds in the region, nor do we expect line-of-sight (LOS)
absorption associated with this high $T_R$.  For spectra B--K, the
best fit to a Gaussian is overlaid and the line width (defined as the
FWHM) in \kms ~is labeled in the upper right-hand corner of each
panel.  The NH\3 (6,6) line widths are high throughout the central
10~pc.  The typical line width for cool molecular gas at the Galactic
center is $\sim15$~\kms (Armstrong \& Barrett 1985; MCH), but none of
the NH\3(6,6) features has a line width of less than 20~\kms, despite
our sensitivity to FWHM$\simgt6$~\kms.

\begin{figure}
\plotone{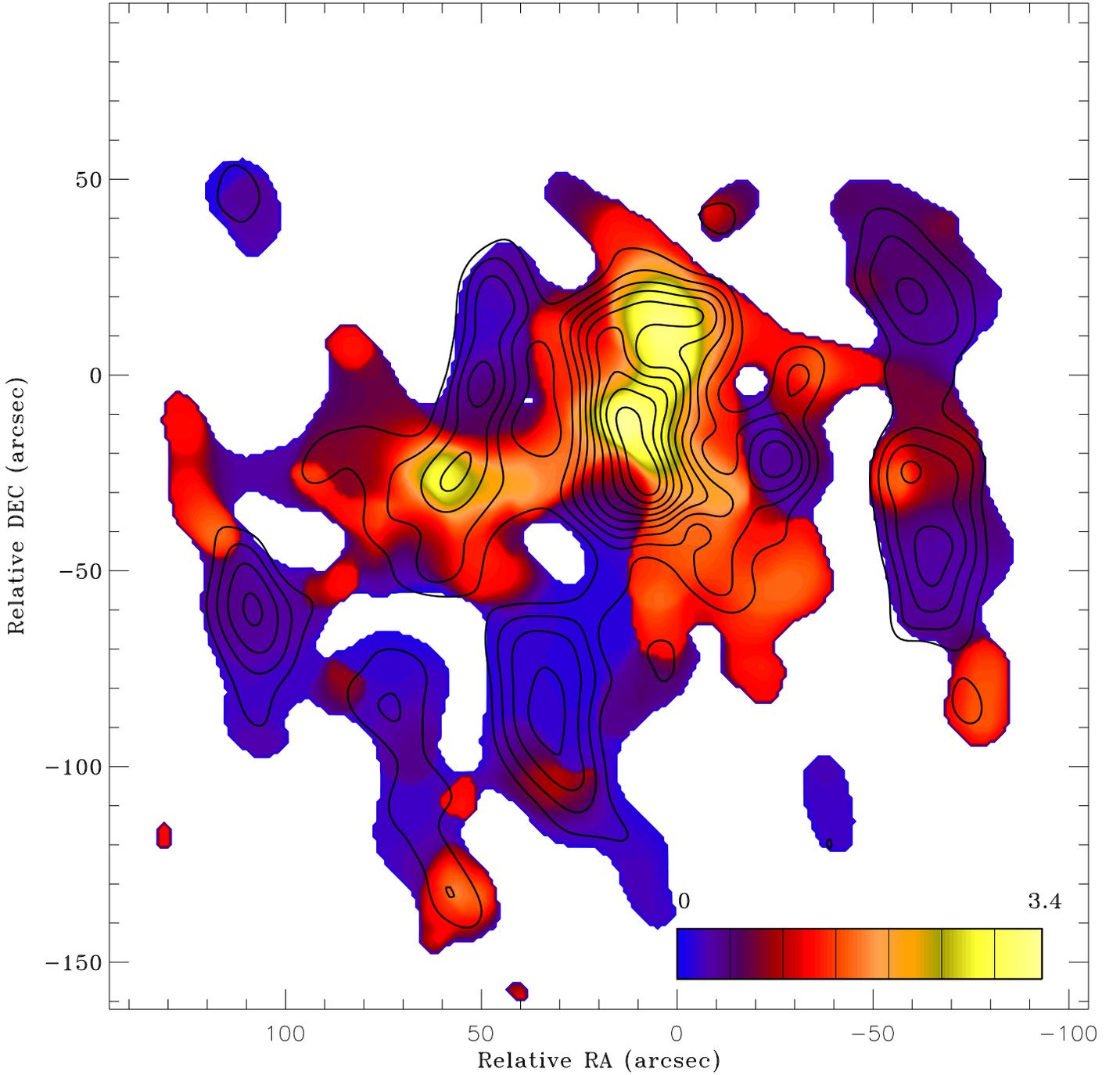}
\caption{Ratio of main line NH\3(6,6) and (3,3) emission. The velocity
channel for the main line is chosen using the NH\3(6,6) image cube.
Line ratios are calculated for every point with (6,6) emission
$>3\sigma'_{ch}$.  For pixels with faint (3,3) emission,
$3\sigma_{33}$ is used to estimate a lower limit for the line
ratio. Sgr A* is located at (0,0). \label{lrat.fig}}
\end{figure}

Finally, we show the line ratio of NH\3(6,6) to (3,3) emission in
color in Figure \ref{lrat.fig}.  To match the resolution of the
NH\3(3,3) image, the NH\3(6,6) image was remade using a Gaussian taper
resulting in $15''\times13''$ resolution and $\sigma'_{ch}=11$~mJy
beam$^{-1}$, which is equal to the noise in the (3,3) image,
$\sigma_{33}$ (MCH).  The velocity channel corresponding to peak
NH\3(6,6) emission is used to measure both the (6,6) and (3,3) flux
densities which are then used to calculate the line ratios.  Only
pixels with NH\3(6,6) flux density greater than $3\sigma'_{ch}$ are
included.  For those pixels where NH\3(3,3) emission is less than
$3\sigma_{33}$, the ratio of (6,6) emission to $3\sigma_{33}$ is used
as a lower limit for the line ratio.  Contours of the velocity
integrated NH\3(6,6) emission are overlaid for reference.

If NH\3(6,6) and (3,3) are in thermal equilibrium, the line ratio
 gives an estimate of the rotational temperature of the gas.
From \citet{tow75}, $T_R$ is related to the opacity of the NH\3(3,3)
main line and the ratio of NH\3(6,6) and (3,3) antenna temperatures by
\bq
\rm{T_{R}}=\frac{-287~\rm{K}}{\rm{ln}[\frac{-0.43}{\tau_m(3,3)}\rm{ln}(1-\frac{\Delta
T_A(6,6)}{\Delta T_A(3,3)}(1-e^{-\tau_m(3,3)}))]} \eq
\noindent if equal beam filling factors are assumed.  
The exponential dependence of $T_R$ on the line ratio reduces the
accuracy of the estimation for $T_R\simgt1000$~K.  This is a
significant improvement on NH\3(2,2) to (1,1) line ratios which only
give accurate estimations for $T_R\simlt50$~K.  The rotational
temperature is thought to be a good indicator of kinetic temperature,
$T_K$, and we assume $T_R$ gives a lower limit for $T_K$
\citep{mar82}.

Line ratios greater than 2.3 are seen in Figure \ref{lrat.fig}, but
they give $T_R<0$.  These large line ratios may the the result of
a larger filling factor for NH\3(6,6), or the dynamic range of the
NH\3(3,3) data may be limited by nearby bright emission (MCH). It is
unlikely that the gas is out of thermal equilibrium because the
equilibration time is only $\sim10^3$~s.  Line widths are quite
large in the region (50 -- 80 \kms) making it unlikely for the
NH\3(6,6) population to be inverted.  The large line widths also make
absorption of NH\3(3,3) by an un-associated, cool foreground cloud
unlikely.  If the NH\3(6,6) originates in a radiatively heated cloud,
then it is possible that the NH\3(3,3) is absorbed by cooler material
in the same cloud that has been shielded from the radiation.  The
NH\3(6,6) would be unaffected by absorption because the cooler gas
would contain almost no NH\3(6,6).  For clouds near the nucleus, the
radiation involved would most likely originate in the central stellar
clusters and those clouds with line ratios $>2.3$ would be located in
front of the nucleus along the LOS.

\subsection{{\it The central 1.5 parsecs}}
The NH\3(6,6) velocity integrated map is dominated by emission less
than 1.5~pc (40$''$) from Sgr A*, interior to the CND (spectra B, C,
\& D).  Indications of molecular gas inside the CND have also been
observed in NH\3(3,3), but the emission was much fainter than
surrounding features (MCH).  Observations of 63~$\mu$m emission from
[O\sc i\rm] with a resolution of $22''$ indicate a central
concentration of neutral gas interior to the CND, possibly associated
with the arcs of Sgr A West \citep{jac93}.  The distribution of the
NH\3(6,6) does not mimic the shape of Sgr A West, and the line
profiles do not show the high velocities associated with the ionized
arcs and [O\sc i\rm] emission \citep{jac93,rob93}.

The NH\3(6,6) emission appears to be kinematically independent of
material in the CND.  To the north of Sgr A* (spectra B and C), the
gas has a velocity of $\sim0$~\kms, but HCN(1-0) and HCO$^+$(1-0)
emission from nearby parts of the CND have velocities of $50-100$~\kms
~\citep{wri01}.  To the southeast of Sgr A* (spectrum D) emission is
coincident with a gap in the CND and possible origins for this cloud
are discussed at the end of this section.  Spectrum E has a velocity
of --20~\kms ~in NH\3(6,6) compared to --5~\kms ~in NH\3(3,3).
HCN(1-0) and HCO$^+$(1-0) spectra at this position are affected by
absorption and central velocities cannot be determined, but emission
from the nearby southern lobe of the CND is centered at $-100$~\kms
~\citep{wri01}.  The NH\3(6,6) line has a relatively small line width
(48 \kms) and low line ratio (0.3) compared to spectra B, C, and D and
we assume this feature is not located in the central parsec.

The large line widths of 75--85~\kms ~in the central 1.5~pc (spectra
B, C, and D) indicate that this gas is physically close to the
nucleus.  The line profiles are similar to those from the CND for
other molecules (e.g. \citet{wri01}) and are well-fitted by Gaussians.
The smooth profiles make turbulence an unlikely broadening mechanism.
Observations of the expansion of SNRs into molecular material show
that the resulting line profiles are often best fit by a combination
of dissociative J-type and non-dissociative C-type shocks
\citep{wan92,van93}.  If the material is traveling perpendicular to
the LOS, the combination of the two shocks results in a symmetrically
broadened profile \citep{van93}.  Thus an expanding shock front
perpendicular to our LOS may explain the large observed line widths.

The NH\3(6,6) to (3,3) line ratios are very high for spectra B, C, and
D and many of the features in the central 1.5~pc exceed the upper
limit for the line ratio of 2.3.  A direct calculation of $T_R$ is not
possible, but the trends in Figure \ref{lrat.fig} indicate that gas
near the nucleus is hot and we assume a warm kinetic temperature of
500~K.  For an approximate mass estimate, we assume an optical depth of
0.1 and $N_{\rm{H}_2}=10^8N_{\rm{NH}_3}$ \citep{ser86} and find
$N_{\rm{H}_2}=8-9\times10^{23}$~cm$^{-2}$ for B, C, and D, and a total
gas mass in the central 2~pc on the order of $10^4$~$M_\odot$.  This
mass is similar to those calculated by \citep{coi99,coi00} for cooler
molecular clouds in the central 10~pc. The estimated virial mass
is $\sim10^6$~$M_\odot$ for a cloud of radius 15$''$ and line width
80~\kms \citep{mac88}.  Thus, it is safe to assume that these clouds
are not gravitationally bound.

In lower transitions of NH\3, there is evidence for a connection of
the ``20~\kms'' GMC to the nucleus along the ``southern streamer''
($\Delta\alpha=30''$, $\Delta\delta=-100''$) \citep{ho91,coi99,coi00}.
The gas appears to become warmer and line widths broaden as the
southern streamer approaches the nucleus.  However, the emission
disappears within 2~pc of the nucleus.  We detect significant
NH\3(6,6) $\sim20''$ to the southeast of Sgr A*, coincident with the
``gap'' in the CND seen in HCN(1-0) and HCO$^+$(1-0) \citep{wri01}.
This gas has a velocity of 40~\kms and appears to be kinematically
connected to the southern streamer (spectrum I).  The large line
ratios and broadened line widths (up to 80~\kms) indicate that it may
trace the continuation of the southern streamer towards the nucleus.
The velocity is less than the escape velocity for a distance of 2~pc,
but ambiguities in projected distance and velocity make it impossible
to claim that it is definitely bound by the nucleus.

\subsection{{\it Features outside the CND}}
The ``western streamer'' is offset $1'$ to the west of Sgr A* along
the edge of Sgr A East (coincident with spectra G \& H).  It is known
to have high ratios of NH\3(2,2) to (1,1) emission (MCH) as well as
large line widths \citep{mcg02}.  In NH\3(6,6), line widths range
between 30 and 50~\kms.  The line ratios in the western streamer have
an average value of 0.66 indicating $T_R\approx230$~K if the gas is
optically thin.  The velocity gradient of 25~\kms~pc$^{-1}$ observed
in NH\3 by MCH accounts for the line velocities of +30~\kms ~in
spectrum G and --30~\kms in spectrum H. The gradient is easily
explained by a long filament highly inclined to the the LOS and moving
out with the expanding shell.  The impact of the shell perpendicular
to the LOS could again explain the broad line widths in this feature.

The molecular ridge that connects the ``20~\kms'' GMC in the south to
the ``50~\kms'' GMC in the east (Ho et al. 1991; Dent et al. 1993;
Coil \& Ho 1999; 2000, MCH) is composed of the southern part of the
southern streamer, SE1 ($70''$, $-100''$), and SE2 ($110''$, $-60''$)
(spectra I, J, \& K, respectively).  Emission from the molecular ridge
 has a typical velocity of 35~\kms, line width of 25~\kms, ~and an
average line ratio of 0.41 (170~K), making it relatively cool
compared to other features in our map.  However, the line widths are
roughly 10~\kms ~larger than that associated with the nearby, cold
GMCs.  The concave southeastern edge of Sgr A East seen in the 6~cm
image is thought to result from an impact with SNR G359.92--0.09
(centered at $\Delta\alpha=120''$, $\Delta\delta=-180''$,
\citep{coi00}) and it is possible that the shocks from this collision
have produced increased temperatures and densities \citep{coi00}.
There is a lack of NH\3(6,6) emission on the northeast edge of Sgr A
East where it is impacting the massive, but relatively quiescent,
``50~\kms'' GMC.  The high density of this GMC may result in a faster
cooling rate in this feature, limiting the production of NH\3(6,6).

Within the outer edge of the Sgr A East shell is a well-known
north-south ridge of continuum radio emission $\sim50''$ to the east
of Sgr A* (coincident with spectrum F), roughly along the eastern edge
of Sgr A West \citep{ped89}.   Although the northern end of the
ridge is coincident with the eastern edge of the CND, NH\3(3,3)
observations indicate that the 50~\kms ~gas is most likely an
extension of SE1 (MCH).  The NH\3(6,6) line width along this ridge
ranges from 40--60~\kms ~and has an average line ratio of 1.0
corresponding to $T_R=340$~K.  The line ratio peaks in the south where
there is a lack of emission from HCN \citep{wri01}.  The similarity of
the western streamer and this eastern ridge as well as the asymmetric
line profile (spectrum F) indicates that it likely lies on the
expanding front of Sgr A East \citep{ped89}.

\section{Summary}
With 25~GHz observations now practical at the VLA, NH\3(6,6) is an
excellent tracer of hot, dense molecular material.  Velocity
integrated NH\3(6,6) images are dominated by emission less than 1.5~pc
in projected distance from Sgr A*.  High line ratios and line widths
indicate that this molecular material is likely close to the
supermassive black hole and is interior to the CND.  In addition, the
hottest emission to the southeast of Sgr A* appears to connect the
southern streamer to the nucleus. The remaining NH\3(6,6) emission
tends to lie along the ridges that surround the Sgr A East shell,
tracing features that show a high line width in earlier NH\3(1,1),
(2,2), and (3,3) data.

\acknowledgements{The authors thank P. Sollins, T. Bergin, and the
 referee for helpful comments and D. Shepherd for help with the 25~GHz
 system.  RSM is supported in part by a Harvard University Merit
 Fellowship.}

\newpage

\newpage

\end{document}